\documentclass[12pt]{amsart}
\usepackage{amsmath,amsthm,amssymb}
\usepackage[margin=1in]{geometry}
\usepackage{hyperref,url}
\numberwithin{equation}{section}

\theoremstyle{plain}
\newtheorem{theorem}{Theorem}[section]
\newtheorem{lemma}[theorem]{Lemma}
\newtheorem{proposition}[theorem]{Proposition}
\newtheorem{corollary}[theorem]{Corollary}

\theoremstyle{definition}
\newtheorem{definition}[theorem]{Definition}

\theoremstyle{remark}
\newtheorem{remark}[theorem]{Remark}

\newcommand{\be}{\begin{equation}}
\newcommand{\ee}{\end{equation}}
\newcommand{\R}{\mathbb{R}}
\newcommand{\M}{\mathcal M}
\newcommand{\B}{\mathcal B}
\newcommand{\E}{\mathcal E}
\newcommand{\K}{\mathcal{K}}
\newcommand{\J}{\mathcal{J}}
\newcommand{\h}{\mathcal{H}}
\newcommand{\U}{\mathcal{U}}

\newcommand{\mg}{\mathfrak{g}}
\newcommand{\mh}{\mathfrak{h}}
\newcommand{\mm}{\mathfrak{m}}

\title{Fundamental Symmetries in Krein Spaces}
\author{Arkadiusz Jadczyk}
\address{Faculty of Physics and Astronomy at the University of Wrocław, plac Maxa Borna 9, 50-204 Wrocław, Poland}
\email{ajadczyk@physics.org}
\begin{document}
\begin{abstract}
We study the space of fundamental symmetries of a real or complex
Krein space in finite or infinite dimension. After recalling the
equivalence of the associated Hilbert norms, we describe two global
operator parametrizations and their relation. The space is realized
as a Banach symmetric space under $m(J,J')=JJ'J$. Its canonical
connection is the tangential projection of the flat derivative in
the real Banach space of Krein-self-adjoint operators. This yields
explicit formulas for geodesics, parallel transport, transvections,
and curvature. The positive square roots $(JJ')^{1/2}$ implement
both isometries between the associated Hilbert spaces and geodesic
transvections. A proof of their intertwining property in an earlier
publication is corrected.
\end{abstract}
\maketitle
\section{Introduction}
\label{sec:introduction}

A fundamental symmetry on a Krein space converts its indefinite
inner product into a Hilbert space inner product. Although this
choice is not unique, the resulting Hilbert norms are equivalent.
The space $\mathcal J(\mathcal K)$ of all fundamental symmetries
therefore provides a natural setting in which to study the relation
between these Hilbert space structures.

This paper gives an operator-theoretic account of
$\mathcal J(\mathcal K)$ and its canonical affine geometry, for both
real and complex Krein spaces of arbitrary dimension. We first
recall the basic norm-equivalence results and describe two classical
global parametrizations relative to a reference symmetry: one by an
unrestricted bounded operator and the other by an angular operator
of norm less than one. These constructions belong to the standard
theory of indefinite inner product spaces
\cite{bognar,azizov,ikl}.

We then realize $\mathcal J(\mathcal K)$ as a Banach symmetric space
with multiplication $m(J,J')=JJ'J$, within the general framework of
\cite{loos,bertram2,klotz}. The canonical connection is expressed as
the projection of the flat derivative in the ambient space of
Krein-self-adjoint operators. This yields explicit formulas for
geodesics, parallel transport, transvections, and curvature, without
requiring a Riemannian metric. In particular, the geodesic joining
$J$ to $J'$ is $s\mapsto(J'J)^sJ$, and the curvature in the
tangent-operator realization is
$R_J(\xi,\eta)\zeta=\frac14[[\xi,\eta],\zeta]$,
with the convention specified below. The square-root operators
$t(J,J')=(JJ')^{1/2}$ connect the Hilbert space and affine-geometric
descriptions. We also correct the proof of their intertwining
property given in \cite{jad72} (Lemma~3.1).

\section{Preliminaries}
\label{sec:preliminaries}

Let $\mathcal{K}$ be a vector space, finite or infinite dimensional, real or complex. We will denote by $(\cdot,\cdot)$ a nondegenerate Hermitian sesquilinear form on $\mathcal{K}$. If $\mathcal{K}$ is real, then ``sesquilinear'' means ``bilinear symmetric''. If $\mathcal{K}$ is complex, we assume that $(x,y)$ is antilinear in $x$, and linear in $y$.
\begin{remark}
In the literature the indefinite scalar product is often written as $[x,y]$. We will use the notation $(x,y)$ as it will cause no conflict with our other notation.
\end{remark}
The symbol $I$ denotes the identity operator on the space indicated by the context.
\begin{definition} A linear operator $J:\mathcal{K}\rightarrow \mathcal{K}$ is called a fundamental symmetry if
\begin{enumerate}
\item[{\rm (i)}] $(Jx,y)=(x,Jy)$ for all $x,y\in \mathcal{K}$,
\item [{\rm (ii)}] $J^2=I,$
\item [{\rm (iii)}] $(x,y)_J:=(x,Jy)$ is a (real or complex) Hilbert space  scalar product on $\mathcal{K}$.
\end{enumerate}
We denote by $\mathcal{J}(\mathcal{K})$ the set of all fundamental symmetries on $\mathcal{K}$.
We call $(\mathcal K,(\cdot,\cdot))$ a Krein space if $\mathcal J(\mathcal K)\ne\varnothing$. We assume throughout that the form is indefinite; thus both eigenspaces of a fundamental symmetry are nonzero.
\end{definition}
\noindent With condition (i), the condition (ii) can be equivalently written as
\begin{equation} (Jx,Jy)=(x,y),\end{equation}
for all $x,y\in\mathcal{K}$. In other words: $J$ is an isometry of $(\K,(\cdot,\cdot))$.

\medskip

If $J\in\mathcal{J}(\mathcal{K})$, we denote by $\mathcal{H}_J$ the space $\mathcal{K}$ endowed with the scalar product $(x,y)_J$, and by $\Vert x \Vert_J$ the corresponding norm:

\[ \Vert x\Vert_J^2=(x,x)_J.\]

Condition (iii) includes positivity and completeness:
$(x,Jx)>0$ for $x\ne0$, and $\mathcal H_J$ is complete in
$\|\cdot\|_J$.
We denote by $L(\mathcal{H}_J)$ the algebra of all bounded linear operators on $\mathcal{H}_J$. For any $A\in L(\mathcal{H}_J)$ we denote by $A^J$ the adjoint of $A$ defined by
\begin{equation} (A^Jx,y)_J=(x,Ay)_J.\end{equation}
We recall now the Hellinger--Toeplitz theorem (see e.g.\ \cite{weidmann} (\S 4.1)).
\begin{theorem}
If $\mathcal{H},\mathcal{H}'$ are Hilbert spaces and $A:\mathcal{H}\rightarrow \mathcal{H}',\, B:\mathcal{H}'\rightarrow \mathcal{H}$ are linear maps satisfying
\[(Ax,z)=(x,Bz)\]
for all $x\in \mathcal{H}$, $z\in \mathcal{H}'$, then $A$ and $B$ are bounded.
\end{theorem}
\noindent The scalar product on the left is in $\mathcal{H}'$, that on the right is in $\mathcal{H}$.

\medskip

If $A=A^J$ we say that $A$ is $J$-self-adjoint.
The first important property of any $J\in\mathcal{J}(\mathcal{K})$ is that it is $J$-self-adjoint.
\begin{proposition}\label{prop:Jselfadjoint}
For every $J\in \mathcal{J}(\mathcal{K})$ we have: $J\in L(\mathcal{H}_J)$ and  $J^J=J.$
\end{proposition}
\begin{proof}
Since $\|Jx\|_J^2=(Jx,x)=(x,Jx)=\|x\|_J^2$, $J$ is bounded. Moreover,
$(Jx,y)_J=(Jx,Jy)=(x,y)=(x,Jy)_J$, so $J^J=J$.
\end{proof}
\begin{definition}If $(\mathcal{H},(\cdot,\cdot))$ is a Hilbert space (real or complex), we call an operator $A\in L(\mathcal{H})$ \emph{positive} if $A$ is self-adjoint and $(x,Ax)\geq 0$ for all $x\in \mathcal{H}$.
\end{definition}
\begin{proposition}\label{prop:JJprimepositive}
If $J,J'$ are in $\mathcal{J}(\mathcal{K})$, then $JJ'$ is bounded positive (therefore self-adjoint) on both $\mathcal{H}_J$ and $\mathcal{H}_{J'}$.
\end{proposition}
\begin{proof}
We have
\[ (x,JJ'y)_J=(x,J^2J'y)=(x,J'y)=(J'x,y)=(J'x,Jy)_J=(JJ'x,y)_J,\]
therefore, by the Hellinger--Toeplitz theorem, $JJ'$ is bounded and self-adjoint on $\mathcal{H}_J$. Similarly
\[ (x,JJ'y)_{J'}=(x,J'JJ'y)=(J'x,JJ'y)=(JJ'x,J'y)=(JJ'x,y)_{J'},\]
therefore $JJ'$ is bounded and self-adjoint on $\mathcal{H}_{J'}$. Furthermore
\[ (x,JJ'x)_J=(x,J'x)=\| x\|^2_{J'}\geq 0.\]
Similarly
\[ (x,JJ'x)_{J'}=(x,J'JJ'x)=(J'x,JJ'x)=\| J'x\|^2_J\geq 0.\]
\end{proof}
\begin{corollary}\label{cor:normequiv}
For any $J,J'\in \mathcal{J}(\mathcal{K})$ the Hilbert space norms for $\mathcal{H}_J$ and $\mathcal{H}_{J'}$ are equivalent.
\end{corollary}
\begin{proof}
By Proposition~\ref{prop:JJprimepositive},
\[
\|x\|_{J'}^2=(x,JJ'x)_J
\le\|JJ'\|_J\|x\|_J^2.
\]
Interchanging $J$ and $J'$ gives the reverse inequality.
\end{proof}
\begin{definition}\label{deflk}It follows that for any two $J,J'\in\mathcal{J}(\mathcal{K})$ the algebras $L(\mathcal{H}_J)$ and $L(\mathcal{H}_{J'})$ coincide. We will use the notation $L(\mathcal{K})$ for this one algebra, and use the term \emph{bounded operator} for elements of $L(\mathcal{K})$.   For any $T\in L(\mathcal{K})$ we denote by $T^*$ its adjoint with respect to the Krein scalar product, so that
\begin{equation} (T^*x,y)=(x,Ty),\quad (x,y\in \mathcal{K}).\end{equation}
For any $J\in \mathcal{J}(\mathcal{K})$ we have then
\begin{equation} A^*=JA^J J,\, A^J=JA^*J.\end{equation}
Evidently $J=J^*$ and $J=E_J^+-E_J^-$, where $E_J^\pm=(I\pm J)/2$, ${E_J^\pm}^2=E_J^\pm,$ ${E_J^\pm}^*={E_J^\pm}^J=E_J^\pm$ are $\mathcal{K}$--, and $\h_J$--orthogonal projections onto $\pm 1$ eigenvalue subspaces of $J$.
\end{definition}

\section{Parametrization of $\mathcal{J}(\mathcal{K})$}

In order to obtain a parametrization of the space of fundamental symmetries we need to select an ``origin'' $J_0$. Then the parametrization will be with respect to this fixed point. Once $J_0$ is fixed, $\mathcal{K}$ decomposes into a direct sum of two Hilbert spaces \begin{equation} \mathcal{K}=\mathcal{H}_0^+\oplus \mathcal{H}_0^-,\label{eqsplit}\end{equation} the eigenspaces of $J_0$ corresponding to eigenvalues $+1$ and $-1$. Corresponding to this decomposition any bounded operator on $\mathcal{K}$ will be written as a block matrix
\be T=\begin{pmatrix}A&B\\C&D\end{pmatrix},\label{eqblock}\ee
with $A\in L(\mathcal{H}_0^+)$, $B\in L(\mathcal{H}_0^-,\mathcal{H}_0^+)$, $C\in L(\mathcal{H}_0^+,\mathcal{H}_0^-)$, $D\in L(\mathcal{H}_0^-)$. Every $x\in \mathcal{K}$ can be written as
\begin{equation} x=\begin{pmatrix}x_+\\x_-\end{pmatrix},\label{eqxpxm}\end{equation}
with $x_\pm\in\mathcal{H}_0^\pm$, and \[(x,y)=(x_+,y_+)_+-(x_-,y_-)_-,\]
\[(x,y)_{J_0}=(x_+,y_+)_++(x_-,y_-)_-,\]
with the evident notation.

\medskip

To simplify the notation we will skip the subscript $J$ in $\mathcal{H}_J^\pm$ and $E_J^\pm$. Let now $J$ be any other fundamental symmetry, and let $\mathcal{K}=\mathcal{H}^+\oplus \mathcal{H}^-$ be the decomposition of $\mathcal{K}$ with $\mathcal{H}^\pm$ the two eigensubspaces of $J$, and $E^\pm$ the corresponding $\mathcal{K}$-orthogonal projections.

\subsection{General form of $J$}

{\bf Notation.} In this section we will use the symbol ${}^*$ to denote Hilbert space adjoints for operators in and between the Hilbert spaces $\mathcal{H}_0^+$ and $\mathcal{H}_0^-$, i.e.\ for the blocks $A,B,C,D$ of the operators written relative to the fixed reference decomposition $\mathcal{K}=\mathcal{H}_0^+\oplus\mathcal{H}_0^-$.

\subsubsection{The unrestricted parametrization}
Here ``unrestricted'' refers to the absence of a norm bound on the
parameter; all operators considered are bounded.

Let $J\in \mathcal{J}(\mathcal{K})$. The condition $(Jx,y)=(x,Jy)$ translates to $(Jx,J_0y)_0=(x,J_0Jy)_0$, or
\begin{equation} J^{J_0}J_0=J_0J.\label{eqjj0}\end{equation}
In what follows we will use one and the same symbol  $I$ to denote the identity operator in $\mathcal{H}_0^\pm$, and in $\K$. Writing $J_0$ and $J$ in block matrix form:
\begin{equation} J_0=\begin{pmatrix}I&0\\0&-I\end{pmatrix},\quad J=\begin{pmatrix}A&B\\C&D\end{pmatrix},\end{equation}
the condition (\ref{eqjj0}) entails
\begin{equation} A=A^*, D=D^*,C=-B^*.\end{equation}
Here the star $*$ stands for the adjoint with respect to the Hilbert space structures in $\h_{J_0}^\pm$. Thus $J$ is of the form
\begin{equation} J=\begin{pmatrix}A&B\\-B^*&D\end{pmatrix},\end{equation}
with $A$ (resp.\ $D$) self-adjoint on $\h_0^+$ (resp. $\h_0^-$).
The additional condition $J^2=I$ results in
\begin{equation} A^2=I+BB^*, \quad D^2=I+B^*B, \quad AB+BD=0.\label{eqj2}\end{equation}
The requirement that the scalar product $(x,y)_J=(x,Jy)=(x,J_0Jy)_0$ be positive-definite entails
\begin{equation} (u,Au)+(u,Bv)+(v,B^*u)-(v,Dv)\geq 0\label{eqineq}\end{equation}
for all $u\in \mathcal{H}_0^+,v\in \mathcal{H}_0^-$, and the result is zero if and only if $u=0$ and $v=0$. In particular, setting $v=0$ we obtain $A\geq 0$, while setting $u=0$ leads to $D\leq 0$. Thus, from (\ref{eqj2}) we find that:
\begin{equation} A=(I+BB^*)^\frac12,\, D=-(I+B^*B)^\frac12.\label{eqAD}\end{equation}
Thus
\begin{equation} J=\begin{pmatrix}(I+BB^*)^\frac12&B\\-B^*&-(I+B^*B)^\frac12\end{pmatrix}.\label{eqJB}\end{equation}
\medskip
\begin{remark}\label{remfx}
We use the identities
\begin{equation}
Xf(X^*X)=f(XX^*)X,\qquad
X^*f(XX^*)=f(X^*X)X^*.
\label{eqident}
\end{equation}
Here $X$ is a bounded operator between Hilbert spaces and $f$ is
continuous on $[0,\|X\|^2]$. The identities hold for polynomials
and hence for continuous functions by uniform polynomial approximation.
\end{remark}
\medskip
Using this identity with $X=B$ and $f(t)=(1+t)^{1/2}$, one checks that $A,D$ given by (\ref{eqAD}) automatically satisfy the off-diagonal relation $AB+BD=0$ in (\ref{eqj2}), so that this relation imposes no further restriction on $B$.
\begin{proposition}[Unrestricted parametrization]\label{prop:unbounded}
Let $J_0\in\mathcal{J}(\mathcal{K})$ be fixed, with associated decomposition $\mathcal{K}=\mathcal{H}_0^+\oplus\mathcal{H}_0^-$. For every $B\in L(\mathcal{H}_0^-,\mathcal{H}_0^+)$ the operator
\[ J_B:=\begin{pmatrix}(I+BB^*)^{1/2}&B\\-B^*&-(I+B^*B)^{1/2}\end{pmatrix}\]
belongs to $\mathcal{J}(\mathcal{K})$. Conversely, every $J\in\mathcal{J}(\mathcal{K})$ is of the form $J=J_B$ for a unique $B\in L(\mathcal{H}_0^-,\mathcal{H}_0^+)$. In particular $J=J_0$ corresponds to $B=0$.
\end{proposition}
\begin{proof}
The preceding calculation proves necessity and uniqueness.
Conversely, for $J$ given by \eqref{eqJB}, Krein self-adjointness
and $J^2=I$ have already been verified. It remains to prove
positivity and completeness. The block matrix of $J_0J$ is
\be J_0J=\begin{pmatrix}A&B\\B^*&D'\end{pmatrix},\ee
where 
\be A=(I+BB^*)^\frac12,\quad  D'=(I+B^*B)^\frac12.\ee
This can be factorized as
\be J_0J=L(B)\begin{pmatrix}A^{-1}&0\\0& D'\end{pmatrix}L(B)^*,\quad L(B)=\begin{pmatrix}I&BD'^{-1}\\0&I\end{pmatrix}.\ee
To verify the factorization one uses the identity $A^{-1}+B(D')^{-1}B^*=A$, which can be seen to hold by using (\ref{eqident}). The diagonal operator is bounded below by a positive constant, and
$L(B)$ is boundedly invertible, with $L(B)^{-1}=L(-B)$.
Thus $J_0J$ is uniformly positive. Its quadratic form defines a norm
equivalent to the reference Hilbert norm and is therefore complete. 
\end{proof}

\subsubsection{The bounded parametrization}

Write $J=J_B$ as in Proposition~\ref{prop:unbounded}, with
$A=(I+BB^*)^{1/2}$, and define
\[
P=-(A+I)^{-1}B
  =-B\bigl(I+(I+B^*B)^{1/2}\bigr)^{-1}.
\]
The equation $Jx=-x$ and the identities in Remark~\ref{remfx} give
\[
\mathcal H^-=
\left\{\binom{Pu}{u}:u\in\mathcal H_0^-\right\}.
\]
Moreover, continuous functional calculus gives
\[
\|P\|=\frac{\|B\|}{1+\sqrt{1+\|B\|^2}}<1.
\]
The operator $P$ is the angular operator of $\mathcal H^-$
\cite{azizov} (Chapter~1, Section~8).
Krein orthogonality then yields
\[
\mathcal H^+=
\left\{\binom{v}{P^*v}:v\in\mathcal H_0^+\right\}.
\]

Conversely, let $P\in L(\mathcal H_0^-,\mathcal H_0^+)$ satisfy
$\|P\|<1$. Every $x=\binom{x_+}{x_-}$ has a unique decomposition
\begin{equation}
x=\binom{v}{P^*v}+\binom{Pu}{u},
\label{eqxpmP^*}
\end{equation}
where
\begin{align}
u&=(I-P^*P)^{-1}(x_--P^*x_+),\label{equu}\\
v&=(I-PP^*)^{-1}(x_+-Px_-).\label{eqvv}
\end{align}
The two graph subspaces are Krein-orthogonal, and their forms satisfy
\[
\left(\binom{v}{P^*v},\binom{v}{P^*v}\right)
\ge (1-\|P\|^2)\|v\|^2,
\qquad
-\left(\binom{Pu}{u},\binom{Pu}{u}\right)
\ge (1-\|P\|^2)\|u\|^2.
\]
Thus the operator equal to $+I$ on the first graph and $-I$ on
the second is a fundamental symmetry: its Hilbert norm is equivalent
to the reference norm, and hence complete.

\begin{proposition}[Bounded parametrization]\label{prop:bounded}
Let $J_0\in\mathcal{J}(\mathcal{K})$ be fixed, with associated decomposition $\mathcal{K}=\mathcal{H}_0^+\oplus\mathcal{H}_0^-$. For every $J\in\mathcal{J}(\mathcal{K})$ there is a unique angular operator $P\in L(\mathcal{H}_0^-,\mathcal{H}_0^+)$, with $\|P\|<1$, such that the $+1$-eigenspace of $J$ is $\mathcal{H}^+=\{\binom{v}{P^*v}:v\in\mathcal{H}_0^+\}$. Conversely, every contraction $P\in L(\mathcal{H}_0^-,\mathcal{H}_0^+)$ with $\|P\|<1$ arises this way from a unique $J\in\mathcal{J}(\mathcal{K})$. In particular $J_0$ corresponds to $P=0$.
\end{proposition}
The action of $J$ on $x$, represented as in (\ref{eqxpmP^*}), amounts to reversing the sign of the negative component $\binom{Pu}{u}$. Thus
\begin{align} (Jx)_+&=v-Pu,\\
(Jx)_-&=P^*v-u.
\end{align}
where $u$ and $v$ are given by (\ref{equu}), (\ref{eqvv}). We obtain
\[ \begin{split}
(Jx)_+&=(I-PP^*)^{-1}x_+-(I-PP^*)^{-1}Px_-\\
&-P(I-P^*P)^{-1}x_- + P(I-P^*P)^{-1}P^*x_+\\
&=\frac{I+PP^*}{I-PP^*}\,x_+-2(I-PP^*)^{-1}Px_-,
\end{split}\]
and similarly
\be (Jx)_-=2P^*(I-PP^*)^{-1}x_+-\frac{I+P^*P}{I-P^*P}x_-,\ee
using the identity of the Remark \ref{remfx} to combine the terms. Since, on the other hand, we have $(Jx)_+=Ax_++Bx_-,\, (Jx)_-=Cx_++Dx_-,$ we immediately get
\be
\begin{split} 
A&=\frac{I+PP^*}{I-PP^*},\\
B&=-2P(I-P^*P)^{-1},\\
C&=2(I-P^*P)^{-1}P^*,\\
D&=-\frac{I+P^*P}{I-P^*P}.
\label{ABCD}\end{split}\ee
The result is in agreement with the formulas in \cite{azizov} (p.~29).
\subsubsection{Relation between the bounded and unbounded parametrization}

\begin{proposition}[Relation between the two parametrizations]\label{prop:relation}
With $J_0, \mathcal{H}_0^\pm$ as above, let $J\in\mathcal{J}(\mathcal{K})$ correspond to $B\in L(\mathcal{H}_0^-,\mathcal{H}_0^+)$ under the unrestricted parametrization of Proposition~\ref{prop:unbounded}, and to $P\in L(\mathcal{H}_0^-,\mathcal{H}_0^+)$, $\|P\|<1$, under the bounded parametrization of Proposition~\ref{prop:bounded}. Then
\begin{equation}
B=-2(I-PP^*)^{-1}P, \qquad P=-B\left(I+(I+B^*B)^{1/2}\right)^{-1}.
\end{equation}
\end{proposition}
\begin{proof}
The first identity follows from \eqref{ABCD}; the second was
obtained in the construction of the angular operator above.
\end{proof}

\subsubsection{Linear-fractional action of the unitary group $\U(\K)$ on $\J(\K)$ in a bounded parametrization}
Denote by $\U(\K)$ the group of all bounded operators $U\in L(\K)$ such that \be UU^*=U^*U=I\label{equni}.\ee We will call $\U(\K)$ the \emph{unitary group} of $\K$. In the real case it would be more appropriate to call it pseudo-orthogonal group, but we will use the term ``unitary" for both cases. It is evident that every element $U\in \U(\K)$ acts smoothly on $\J(\K)$ by
\be U\in \U(\K): J\in \J(\K)\mapsto U(J)=UJU^*\in \J(\K).\ee
Therefore $\J(\K)$ is a conjugation-invariant submanifold of $\U(\K)$. 
It is also clear that, for each $J$,  $U$ maps, isometrically,  $\h_J^\pm$ onto $\h_{U(J)}^\pm$. If we select a base point $J_0$ in $\J(\K)$, we can represent every $U\in \U(\K)$ as a block matrix as in (\ref{eqblock}). For the blocks, the star denotes the Hilbert adjoint.
Krein unitarity is equivalent to the two sets of identities
\begin{align}
AA^*-BB^*&=I,& DD^*-CC^*&=I,& CA^*-DB^*&=0,\notag\\
A^*A-C^*C&=I,& D^*D-B^*B&=I,& A^*B-C^*D&=0.
\label{eq:block-unitarity}
\end{align}
With respect to the splitting (\ref{eqsplit}) the subspace $\h_J^-$ is represented by
\be x^-=\left\{\begin{pmatrix}Pu\\u\end{pmatrix}:\,u\in \h_0^-\right\},\ee
where $P\in L(\h_0^-,\h_0^+)$ is the bounded representative of $J$. Corresponding to the action $J\mapsto UJU^*$ of $\U(\K)$ on $\J(\K)$ we have action $x^-\mapsto {x'}^-$ which maps $\h_J^-$ onto $\h_{J'}^-$. Now, $\h_{J'}^-$ must be of the same form as $\h_J^-$, but with $P$ replaced by $P'$. With $U$ represented in the block matrix form we obtain
\be Ux_J^-=\begin{pmatrix}A&B\\C&D\end{pmatrix}\begin{pmatrix}Pu\\u\end{pmatrix}
=\begin{pmatrix}P'u'\\u'\end{pmatrix}.\ee
Thus
\begin{align*}
(AP+B)u&=P'u',\\
(CP+D)u&=u'.
\end{align*}
The map $U$ restricts to an isomorphism from $\mathcal H_J^-$
onto $\mathcal H_{J'}^-$. The graph projections of these spaces
onto $\mathcal H_0^-$ are bounded isomorphisms, so $CP+D$ is
boundedly invertible. Consequently,
\be P'=(AP+B)(CP+D)^{-1}.\ee 
\section{Fundamental Symmetries of a Krein Space as a Banach Symmetric Space}
We start with the definition of a symmetric space as given in \cite{loos}. Here we restrict ourselves to the case of a real or complex field and Banach manifolds. 
\begin{definition}A symmetric space is a manifold $\M$ with a differentiable multiplication $m:\,\M\times \M\rightarrow \M$ such that for all $x,y,z\in \M$: 
\begin{enumerate}
\item[{(M1)}] $m(x,x)=x$,
\item[{(M2)}] $m(x,m(x,y))=y$,
\item[{(M3)}] $m(x,m(y,z))=m(m(x,y),m(x,z))$.
\end{enumerate}
The fourth condition exists in two versions. We will use the strong version, as in \cite{loos}. For another version and generalizations - see \cite{bertram, bertram2, klotz}.
\begin{enumerate}
\item[{(M4)}] Every $x$ in $\M$ has a neighborhood such that $m(x,y)=y$ implies $y=x$ for all $y$ in this neighborhood.
\end{enumerate}
\label{defsym}\end{definition}
Any Banach Lie group $G$  can be naturally endowed with the structure of a symmetric space using the multiplication map (see e.g. \cite{loos,klotz}):

\be m(x,y)=xy^{-1}x,\quad (x,y\in G).\label{eqxyx}\ee
The unrestricted parametrization is real-analytic, with inverse
given by extraction of the upper-right block. We use its induced
real Banach manifold structure on $\mathcal J(\mathcal K)$.
The following multiplication makes it a symmetric space.
\begin{proposition}
The formula
\be m(J,J')=JJ'J\label{eqjjj}\ee
defines a multiplication map on $\J(\K)$ satisfying (M1)-(M4), and thus endows $\J(\K)$ with the structure of a symmetric space.
\end{proposition}
\begin{proof}
Since $J$ is Krein-unitary, conjugation by $J$ preserves
$\mathcal J(\mathcal K)$. Smoothness and (M1)--(M3) follow
from the operator formula. To prove (M4), suppose $JJ'J=J'$.
Then $J$ and $J'$ commute, so $Q=JJ'$ satisfies $Q^2=I$.
As $Q$ is positive on $\mathcal H_J$, its spectrum is contained
in $(0,\infty)$; hence $Q=I$ by the spectral theorem.
Thus $J'=J$. In fact, each symmetry has a unique fixed point
on all of $\mathcal J(\mathcal K)$.
\end{proof}
In particular the symmetries $S_J$ of $\M$ are given by
\be S_J(J')=m(J,J') = JJ'J.\label{eqsj}\ee
\subsection{Canonical transvections}
For $J,J'\in\mathcal J(\mathcal K)$, the operator $JJ'$ is
positive and invertible on both $\mathcal H_J$ and $\mathcal H_{J'}$.
Define its positive square root by continuous functional calculus:
\begin{equation}
t(J,J')=(JJ')^{1/2}.
\label{eqtjj}
\end{equation}
The two Hilbert space constructions give the same operator:
continuous functional calculus can be obtained by polynomial
approximation, and the operator norms are equivalent.
The following properties were stated in \cite{jad72} (Lemma~3.1).
The proof below corrects the argument for the intertwining property
in part~(iii) of that lemma.

\begin{proposition}\label{propct}
Let $T=t(J,J')$. Then:
\begin{enumerate}
\item[(i)] $T$ is positive on both $\mathcal H_J$ and $\mathcal H_{J'}$.
\item[(ii)] $T$ is Krein-unitary and
\[
T^*=T^{-1}=t(J',J).
\]
\item[(iii)] $T$ commutes with every bounded operator commuting with
$J$ and $J'$, and
\begin{equation}
TJ'T^{-1}=J.
\label{eq:iii}
\end{equation}
\item[(iv)] The spectrum of $T$ is invariant under
$\lambda\mapsto\lambda^{-1}$. In the real case, spectrum means
the spectrum of the complexified operator.
\item[(v)] The four quantities
\[
\|T\|_J^2,\quad \|T\|_{J'}^2,\quad
\|T^{-1}\|_J^2,\quad \|T^{-1}\|_{J'}^2
\]
are equal, with common value
\[
\rho(JJ')
=\sup_{x\ne0}\frac{\|x\|_{J'}^2}{\|x\|_J^2}
=\sup_{x\ne0}\frac{\|x\|_J^2}{\|x\|_{J'}^2}.
\]
\item[(vi)] $T:\mathcal H_{J'}\to\mathcal H_J$ is a surjective
Hilbert space isometry; in particular, it maps orthonormal bases
to orthonormal bases.
\end{enumerate}
\end{proposition}

\begin{proof}
Put $Q=JJ'$. Since $JQJ=Q^{-1}$, functional calculus gives
\[
JTJ=T^{-1}.
\]
As $T^J=T$, it follows that
\[
T^*=JT^JJ=JTJ=T^{-1}=(J'J)^{1/2}.
\]
Likewise $J'TJ'=T^{-1}$, whence
\[
TJ'T^{-1}=T^2J'=JJ'J'=J.
\]
Commutation with operators commuting with $J,J'$ follows from
polynomial approximation. The similarity $JTJ=T^{-1}$ proves
(iv). Positivity gives
\[
\|T\|_J^2=\|T\|_{J'}^2=\rho(Q),
\qquad
\|T^{-1}\|_J^2=\|T^{-1}\|_{J'}^2=\rho(Q^{-1}).
\]
Since $JQJ=Q^{-1}$, the spectral radii coincide. Moreover,
\[
\|Q\|_J
=\sup_{x\ne0}\frac{(x,Qx)_J}{\|x\|_J^2}
=\sup_{x\ne0}\frac{\|x\|_{J'}^2}{\|x\|_J^2};
\]
interchanging $J,J'$ gives the other quotient. Finally,
$T^{-1}JT=J'$, so
\[
(Tx,Ty)_J=(x,T^*JTy)=(x,J'y)=(x,y)_{J'}.
\]
Invertibility proves surjectivity and the assertion about bases.
\end{proof}

\section{Canonical connection}
{\bf Notation}. To simplify the notation, in the following we will set $\M:=\J(\K)$, $G=\U(\K)$.
 \subsection{The tangent bundle}
 Let $J(t)$ be a smooth  path in $\M$, with $J(0)=J$. Then we have $J(t)=J(t)^*$ and $J(t)^2=I.$ Taking the derivative with respect to $t$ at $t=0$, and denoting the tangent vector at $t=0$ by $\xi$, we obtain
 \be \xi^*=\xi,\, \xi J+J\xi=0.\label{eqxi}\ee
 Conversely, if $\xi\in L(\K)$ satisfies (\ref{eqxi}),put
 \[ X=\frac12\xi J.\]
 Then
 \[X^*=\frac12 J\xi=-\frac12 \xi J=-X,\quad XJ=-JX.\]
 Consequently,
 \[ \gamma(t)=e^{tX}Je^{-tX}\in \M,\]
 and $\dot{\gamma}(0)=[X,J]=\xi.$
 Thus we identify the tangent space at $J$ with the set of all such operators:
 \be T_J\M=\{\xi\in L(\K):\, \xi=\xi^*,\, \xi J+J\xi=0\}.\label{eqtjm}\ee
 \subsection{$\M$ is a reductive homogeneous space}
 It follows, in particular, from Proposition \ref{propct}, that the action of $G$ on $\M$
 \be J\mapsto U(J) = UJU^{-1}\ee
is transitive. The Banach Lie algebra $\mg$ of $G$ is naturally identified with bounded $\K$--anti-Hermitian operators:
 \be \mg=\{X\in L(\K):X=-X^*\}.\ee
\begin{remark}
All manifolds, tangent spaces, and Lie algebras used below are real,
including when $\mathcal K$ is complex. In the complex case the
formula $\mathfrak i_J\xi=iJ\xi$ defines an almost complex structure
on the tangent bundle. Its additional geometric properties are not
needed here.
\end{remark}
At each point $J$ we have a reductive decomposition of $\mg$ into two closed subspaces:
\be \mg=\mh_J\oplus\mm_J,\label{eqred}\ee
where
\be \mh_J=\{X\in L(\K): X=-X^*, XJ=JX\},\quad \mm_J=\{X\in L(\K): X=-X^*, XJ=-JX\}.\ee
We notice that
\be [\mh_J,\mh_J]\subset \mh_J,\, [\mh_J,\mm_J]\subset \mm_J,\,[\mm_J,\mm_J]\subset \mh_J.\ee
Let
\be H_J=\{U\in G: UJU^{-1}=J\}=\{U\in G: UJ=JU\}\ee
be the stability subgroup of $J$. Then 
\be  H_J\sim \U(\h_J^+)\times \U(\h_J^-)\label{eqhxh},\ee
\[ \M\sim G/H_J,\]
$\mh_J$ is the Lie algebra of $H_J$,
and 
\[ \operatorname{Ad}(H_J)\mm_J=\mm_J.\] 
We notice that
\[ J'\mapsto t(J',J)\]
is a smooth section of the orbit map $U\mapsto UJU^{-1}$. Smoothness follows from smooth functional calculus for the square root on the self--adjoint positive invertible operators.
The decomposition (\ref{eqred}) is given by
\be X=\frac12(X+JXJ)+\frac12(X-JXJ),\ee
and the projections of $\mg$ on $\mh_J$ and on $\mm_J$ are evidently continuous. Therefore we can apply all the machinery of Banach reductive homogeneous spaces to our case. In particular each tangent space $T_J$ at $J\in \M$ is naturally identified with $\mm_J$ as follows. Every element $X\in \mg$ generates a {\em fundamental vector field}, denoted $\tilde{X}$, with
\be \tilde{X}(J)=\frac{d}{dt}\left(e^{tX}Je^{-tX}\right)|_{t=0}=[X,J].\label{eqtxj}\ee
In particular, for $X\in \mm_J$ we obtain
\be \tilde{X}(J)=2XJ.\ee
Thus we have the identification $T_J\M\longrightarrow \mm_J$ given by 
\be \xi\in T_J\M\mapsto X_\xi=\frac12 \xi J\in \mm_J.\label{eqXt}\ee
The fundamental vector field $\tilde{X_\xi}$ is then a natural extension of $\xi$ to a vector field on $\M$.

In \cite{bertram3} (p.6) and in \cite{bertram2} (p. 132) the map $\xi\mapsto \tilde{X_\xi}$ is denoted $l_J$. We shall use the following result from these references (\cite{bertram3} (Proposition 1.2.1),\cite{bertram2} (Corollary 26.7)):
\begin{proposition}The covariant derivative associated to the torsion-free canonical connection of a symmetric space $\M$ is given by the formula
\be (\nabla_X Y)_p=[l_p(X_p),Y]_p,\label{eqnab}\ee
where $X,Y$ are arbitrary smooth vector fields on $\M$.\label{propbertram}
\end{proposition}
\begin{remark}The commutator in (\ref{eqnab}) is the commutator of vector fields on a manifold. In particular, if $X,Y$ are in $\mg$, if $Z=[X,Y]$ in $\mg$, and if $\tilde{X},\tilde{Y},\tilde{Z}$ are the corresponding fundamental vector fields on $\M$, then $[\tilde{X},\tilde{Y}]=-\tilde{Z}.$
\end{remark}

\subsection{Projection form of the connection and the curvature}
We express the canonical connection as a projection of the flat
ambient derivative and use this form to compute its curvature.
\label{subsec:projection-curvature}

\noindent{\bf Notation.}
With $\M=\mathcal J(\mathcal K)$ we consider
\[
\E=\{A\in L(\mathcal K):A^*=A\}
\]
as a real Banach space, also when $\mathcal K$ is complex.
The star denotes the Krein adjoint. We use the realization
\[
T_J\M=\{\xi\in \E:\xi J+J\xi=0\}.
\]
For a smooth map $F:\M\to \E$, its differential is denoted by
\[
DF(J):T_J\M\longrightarrow \E.
\]
In particular, a tangent vector field $Y$ is regarded as an
$\E$-valued map, and
\[
D_XY(J):=DY(J)[X(J)].
\]
The operator $D$ is the flat connection on $\M\times \E$.
Our bracket conventions are
\[
[A,B]=AB-BA,
\qquad
[X,Y]=D_XY-D_YX.
\]
The vector-field bracket formula follows from the derivation
interpretation of vector fields; see \cite{schmeding} (Appendix~D, D.8--D.11).
Flatness of $D$ is the identity
$D_XD_Y-D_YD_X=D_{[X,Y]}$.
\begin{definition}
For $J\in \M$, define
\begin{equation}
P_J(A)=\frac12(A-JAJ),
\qquad A\in \E.
\label{eq:pc-projection}
\end{equation}
\end{definition}

\begin{lemma}
The map $P_J$ is a bounded projection with
\[
\operatorname{ran}P_J=T_J\M,
\qquad
\ker P_J=N_J:=\{A\in \E:AJ=JA\},
\]
and
\[
\E=T_J\M\oplus N_J.
\]
For $\xi\in T_J\M$, let $P_\xi=DP(J)[\xi]\in L(\E)$.
Then
\begin{equation}
P_\xi(A)=-\frac12(\xi AJ+JA\xi),
\qquad
P_JP_\xi P_J=0.
\label{eq:pc-projection-identities}
\end{equation}
\end{lemma}

\begin{proof}
The map $A\mapsto JAJ$ is a bounded involution of $\E$.
Its $-1$ and $+1$ eigenspaces are $T_J\M$ and $N_J$,
respectively. Differentiating \eqref{eq:pc-projection} (with respect to the variable $J$)
gives the formula for $P_\xi$. Differentiating $P_J^2=P_J$
and multiplying on both sides by $P_J$ gives
$P_JP_\xi P_J=0$.
\end{proof}

\begin{proposition}
The canonical connection is the tangential projection
of the flat derivative:
\begin{equation}
\begin{aligned}
(\nabla_\xi Y)_J
&=P_J\bigl(DY(J)[\xi]\bigr)\\
&=DY(J)[\xi]
  -\left[\frac12\xi J,Y(J)\right]\\
&=DY(J)[\xi]
  +\frac12\bigl(\xi Y(J)+Y(J)\xi\bigr)J.
\end{aligned}
\label{eq:pc-connection}
\end{equation}
\end{proposition}

\begin{proof}
Fix $J$ and $\xi\in T_J\M$, and set $a=\frac12\xi J$.
The canonical extension of $\xi$ is
\[
l_J(\xi)(Q)=[a,Q],
\qquad Q\in \M.
\]
Since $a$ is fixed,
\[
l_J(\xi)(J)=\xi,
\qquad
D(l_J(\xi))(J)[\eta]=[a,\eta].
\]
The intrinsic formula
$(\nabla_\xi Y)_J=[l_J(\xi),Y](J)$ therefore gives
\[
(\nabla_\xi Y)_J=DY(J)[\xi]-[a,Y(J)].
\]
As $Y(J)J=-JY(J)$,
\[
-[a,Y(J)]
=\frac12\bigl(\xi Y(J)+Y(J)\xi\bigr)J.
\]

To identify this expression with the projection, set
$A=DY(J)[\xi]$. Differentiating
$Y(Q)Q+QY(Q)=0$ at $Q=J$ gives
\[
AJ+JA+Y(J)\xi+\xi Y(J)=0.
\]
Multiplication on the right by $J$ yields
\[
A+JAJ=-\bigl(\xi Y(J)+Y(J)\xi\bigr)J.
\]
Hence
\[
P_J(A)
=A+\frac12\bigl(\xi Y(J)+Y(J)\xi\bigr)J,
\]
which proves \eqref{eq:pc-connection}.
\end{proof}

\begin{corollary}
For a smooth path $J:I\to \M$ and a smooth tangent field
$V$ along $J$,
\begin{equation}
\nabla_sV
=P_{J(s)}(\dot V)
=\dot V+\frac12(\dot JV+V\dot J)J.
\label{eq:pc-along-path}
\end{equation}
In particular, the geodesic equation is
\be 
\ddot J+\dot J^{\,2}J=0.\label{eqgeoex}\ee
\end{corollary}

\begin{proof}
Differentiate $VJ+JV=0$ and apply the preceding calculation.
For the last assertion, take $V=\dot J$.
\end{proof}
\subsubsection{Geodesics}
\begin{proposition}
Let $J\in\mathcal M$ and $X\in\mathfrak m_J$. Then
\begin{equation}
\gamma(s)=e^{sX}Je^{-sX}=e^{2sX}J,\qquad s\in\mathbb R,
\label{eqgeos}
\end{equation}
is a geodesic, and every geodesic is a restriction of such a curve.
For $J,J'\in\mathcal M$, the unique geodesic with
$\gamma(0)=J$ and $\gamma(1)=J'$ is
\[
\gamma(s)=(J'J)^sJ,
\qquad X=\tfrac12\log(J'J).
\]
In particular, the canonical connection is geodesically complete.
\end{proposition}

\begin{proof}
Since $X^*=-X$, $e^{sX}\in G$. Anticommutation gives
$X\gamma(s)=-\gamma(s)X$, and hence
\[
\dot\gamma=2X\gamma,\qquad
\ddot\gamma=4X^2\gamma=-\dot\gamma^{\,2}\gamma.
\]
Thus \eqref{eqgeoex} holds. For any initial tangent vector $\xi$,
the choice $X=\frac12\xi J$ gives $\dot\gamma(0)=\xi$.
Uniqueness for the geodesic initial-value problem proves the
assertion about all geodesics and their extension to $\mathbb R$.

The operator $J'J$ is positive and invertible on $\mathcal H_J$.
Since $J(J'J)J=(J'J)^{-1}$, functional calculus gives
$JXJ=-X$ for $X=\frac12\log(J'J)$. Also $X^J=X$, so
$X^*=JXJ=-X$ and $X\in\mathfrak m_J$. The endpoint identities
follow from $e^{2X}=J'J$.
Conversely, every $X\in\mathfrak m_J$ is self-adjoint on
$\mathcal H_J$, so $\log(e^{2X})=2X$; the endpoint therefore
determines $X$ uniquely.
\end{proof}

\subsubsection{Transvections}
We now recall the definition of a transvection (cf. e.g. \cite{poor} (p. 229))
\begin{definition}
Let $\M$ be a manifold endowed with an affine connection, let $\gamma:\R\longrightarrow \M$ be an affinely parametrized geodesic, and write
\be P_{s,t}^\gamma:T_{\gamma(s)}\M\longrightarrow T_{\gamma(t)}\M\ee
for parallel transport along $\gamma$. A transvection along $\gamma$, with displacement $a$ is an affine map (i.e. a connection preserving diffeomorphism) $\tau_a:\M\longrightarrow \M$ satisfying
\begin{itemize}
\item[{\rm (i)}] $\tau_a(\gamma(s))=\gamma(s+a)$,
\item[{\rm (ii)}] $(d\tau_a)_{\gamma(s)}=P_{s,s+a}^\gamma$ for every $s$.
\end{itemize}
\end{definition}
\begin{proposition}
For a geodesic $\gamma(s)=e^{sX}Je^{-sX}$, the operator
\[
t(\gamma(s+a),\gamma(s))=t(\gamma(a),\gamma(0))=e^{aX}
\]
implements the transvection
$\tau_a(Q)=e^{aX}Qe^{-aX}$. Parallel transport is
\begin{equation}
P^\gamma_{s,s+a}(V)=e^{aX}Ve^{-aX}.
\label{eqgeotr}
\end{equation}
\end{proposition}

\begin{proof}
Equation~\eqref{eq:pc-along-path}, together with
$\dot\gamma=2X\gamma$, reduces $\nabla_sV=0$ to
$\dot V=[X,V]$. This proves \eqref{eqgeotr}. Moreover,
\[
\gamma(s+a)\gamma(s)=e^{2aX}.
\]
As $X^*= -X$ and $X\gamma(s)=-\gamma(s)X$, $X$ is
self-adjoint on $\mathcal H_{\gamma(s)}$.
Thus the positive square root of $e^{2aX}$ is $e^{aX}$.
Finally,
\[
\tau_a=S_{\gamma(a/2)}\circ S_{\gamma(0)},
\]
so $\tau_a$ preserves the canonical connection. Its action on
$\gamma$ is parameter translation, and its differential is
\eqref{eqgeotr}. These are the defining properties of a transvection.
\end{proof}

\begin{remark}
For the general relation between transvections and parallel
transport in Banach symmetric spaces, see
\cite{kim} (Theorem~2.5(vi)).
\end{remark}

\subsubsection{Curvature}
\begin{definition}
We use the curvature convention
\begin{equation}
R(X,Y)Z
=\nabla_X\nabla_YZ-\nabla_Y\nabla_XZ-\nabla_{[X,Y]}Z.
\label{eq:pc-curvature-convention}
\end{equation}
\end{definition}

\begin{lemma}
At $J\in \M$, write $P=P_J$ and
$P_\xi=DP(J)[\xi]$. Then
\begin{equation}
R_J(\xi,\eta)\zeta
=P(P_\xi P_\eta-P_\eta P_\xi)\zeta,
\qquad \xi,\eta,\zeta\in T_J\M.
\label{eq:pc-projection-curvature}
\end{equation}
\end{lemma}

\begin{proof}
Let $X,Y,Z$ extend $\xi,\eta,\zeta$. Write
$P_X(J)=DP(J)[X(J)]$. From $\nabla_XZ=PD_XZ$,
\[
\nabla_X\nabla_YZ
=PP_XD_YZ+PD_X(D_YZ).
\]
The flatness identity
\[
D_X(D_YZ)-D_Y(D_XZ)-D_{[X,Y]}Z=0
\]
therefore gives
\[
R(X,Y)Z=PP_XD_YZ-PP_YD_XZ.
\]
Differentiating $Z=PZ$ gives
\[
D_YZ=P_YZ+PD_YZ.
\]
Since $PP_XP=0$, it follows that
\[
PP_XD_YZ=PP_XP_YZ.
\]
Interchanging $X$ and $Y$ proves the assertion.
\end{proof}

\begin{proposition}
For $\xi,\eta,\zeta\in T_J\M$,
\begin{equation}
R_J(\xi,\eta)\zeta
=\frac14[[\xi,\eta],\zeta].
\label{eq:pc-explicit-curvature}
\end{equation}
\end{proposition}

\begin{proof}
Write $\{A,B\}=AB+BA$. By
\eqref{eq:pc-projection-identities} and the tangent
anticommutation relations,
\[
P_\eta(\zeta)=-\frac12\{\eta,\zeta\}J.
\]
The operator $\{\eta,\zeta\}$ commutes with $J$, and direct
substitution gives
\[
P_\xi P_\eta(\zeta)
=\frac14\{\xi,\{\eta,\zeta\}\}.
\]
Consequently,
\[
\begin{aligned}
(P_\xi P_\eta-P_\eta P_\xi)\zeta
&=\frac14\bigl(
 \{\xi,\{\eta,\zeta\}\}
 -\{\eta,\{\xi,\zeta\}\}\bigr)\\
&=\frac14[[\xi,\eta],\zeta].
\end{aligned}
\]
The last operator is Krein-self-adjoint and anticommutes
with $J$, hence belongs to $T_J\M$. Thus the outer projection
in \eqref{eq:pc-projection-curvature} acts as the identity.
\end{proof}

\begin{remark}
The projection $P_J$ is taken along the commuting
complement $N_J$; no inner product on $\E$ is required.
With the opposite curvature convention, the sign in
\eqref{eq:pc-explicit-curvature} is reversed.
\end{remark}
%\section*{Funding}The author acknowledges support of Quantum Group Inc. 
\section*{Acknowledgements} The author acknowledges support of Quantum Group Inc.  I thank Ricardo Pinho for constructive comments. I am deeply grateful to my wife, Laura Knight-Jadczyk, for her constant support and encouragement. 
%\section*{Conflicts of Interest}The author declares no conflicts of interest.

% Before submission, supply the affiliation and correspondence details,
% and the journal's applicable back-matter declarations: Funding,
% Data Availability, Conflicts of Interest, and (where applicable)
% Author Contributions and Acknowledgments. Confirm all factual statements.
% Disclose substantive generative-AI assistance in accordance with the
% current journal policy; superficial copy-editing alone is exempt.

\end{document}